\documentclass[iop,apj]{emulateapj}
\usepackage{apjfonts}
\usepackage{multirow}
\usepackage{natbib}

\def\aj{{AJ}}

\def\apj{{ApJ}}
\def\apjl{{ApJL}}
\def\apjs{{ApJS}}

\def\lax{{$\mathrel{\hbox{\rlap{\hbox{\lower4pt\hbox{$\sim$}}}\hbox{$<$}}}$}}
\def\gax{{$\mathrel{\hbox{\rlap{\hbox{\lower4pt\hbox{$\sim$}}}\hbox{$>$}}}$}}
\def\simlt{\lower.5ex\hbox{$\; \buildrel < \over \sim \;$}}
\def\simgt{\lower.5ex\hbox{$\; \buildrel > \over \sim \;$}}

\def\mnras{{MNRAS}}
\def\nat{{Natur}}

\def\percm2{cm$^{-2}$}

\def\solmass{$M_\odot$}

\shorttitle{KMTNet Galaxy Survey}
\shortauthors{BYUN ET AL.}

\begin{document}

\title{KMTNet Nearby Galaxy Survey I. : Optimal strategy for low surface brightness imaging with KMTNet}

\author{Woowon Byun\altaffilmark{1,2},
Yun-Kyeong Sheen\altaffilmark{1},
Luis C. Ho\altaffilmark{3,4},
Joon Hyeop Lee\altaffilmark{1,2},
Sang Chul Kim\altaffilmark{1,2},
Hyunjin Jeong\altaffilmark{1},
Byeong-Gon Park\altaffilmark{1,2},
Kwang-Il Seon\altaffilmark{1,2},
Yongseok Lee\altaffilmark{1,5},
Sang-Mok Cha\altaffilmark{1,5},
Minjin Kim\altaffilmark{1,2,6}
}

\altaffiltext{1}{Korea Astronomy and Space Science Institute, Daejeon 34055, Korea}
\altaffiltext{2}{Korea University of Science and Technology, Daejeon 34113, Korea}
\altaffiltext{3}{Kavli Institute for Astronomy and Astrophysics, Peking University, Beijing 100871, China}
\altaffiltext{4}{Department of Astronomy, School of Physics, Peking University, Beijing 100871, China}
\altaffiltext{5}{School of Space Research, Kyung Hee University, Yongin, Kyeonggi 17104, Korea}
\altaffiltext{6}{Department of Astronomy and Atmospheric Sciences, Kyungpook National University, Daegu  702-701, Korea}

\begin{abstract}
In hierarchical galaxy formation models, galaxies evolve through mergers
and accretions. Tidally-disrupted debris from these processes can remain as diffuse, 
faint structures, which can provide useful insight into the assembly 
history of galaxies. To investigate the properties of the faint structures 
in outskirts of nearby galaxies, we conduct deep and wide-field imaging survey 
with KMTNet. We present our observing strategy and optimal data reduction 
process to recover the faint extended features in the imaging data
of NGC 1291 taken with KMTNet. Through the dark sky flat-fielding
and optimal sky subtraction, we can effectively remove inhomogeneous patterns. 
In the combined images, the peak-to-peak global sky 
gradients were reduced to less than $\sim0.5$\% and $\sim0.3$\% of the 
original $B$- and $R$-band sky levels, respectively. 
However, we find local spatial fluctuations in the background sky which can 
affect the precise measurement of the sky value.
Consequently, we can reach the surface brightness of $\mu_{B,1\sigma} \sim$ 
29.5 and $\mu_{R,1\sigma}\sim$ 28.5 mag arcsec$^{-2}$ in azimuthally averaged 
one-dimensional surface brightness profiles, that is mainly limited by the 
uncertainty in the sky determination. 
These results suggest that the deep imaging data produced by KMTNet are 
suitable to study the faint features of nearby galaxies such as 
outer disks and dwarf companions, but unideal (not impossible) to detect 
stellar halos. The one-dimensional profile revealed that
NGC 1291 appeared to have Type I disk out to $R$ $\sim$ 30 kpc with 
no obvious color gradient and excess light due to a stellar halo was undetected.
\end{abstract}


\section{Introduction} 

In the $\Lambda$CDM model, galaxies grow their masses through mergers 
and accretions \citep{1978MNRAS.183..341W}. These processes can leave the 
disturbed structures in the outskirts of galaxies which are remnants 
of tidally-disrupted satellite galaxies \citep{2001ApJ...548...33B,2005ApJ...635..931B,
2008ApJ...689..936J,2010MNRAS.406..744C}. 
While distinct structures such as shells, tails, and stellar streams remain 
from recent or ongoing merger events (t$_{\rm lookback}\sim4-5$ Gyr), 
more ancient mergers form well-mixed stellar halos which have 
extremely low surface brightness (cf. \citealp{2008MNRAS.391...14D,
2008ApJ...689..936J,2018arXiv180403330M}). 
Therefore, the presence and properties of the merging 
features in the galaxy outskirts provide direct hints about 
the recent mass assembly histories of galaxies 
(e.g., \citealp{2001ApJ...548...33B,2008MNRAS.391...14D,2008ApJ...689..936J}).

However, there have not been many studies to observe these signatures 
because of their low surface brightness, followed by expensive 
observational costs. It is generally required to reach a surface 
brightness of $\sim 27-28$ mag arcsec$^{-2}$ to detect tidal debris 
and stellar halos \citep{2011ApJ...739...20C,2012arXiv1204.3082B}. Nevertheless, there have 
been several studies which show the presence of post-merger signatures 
in nearby galaxies. For example, \citet{1979Natur.277..279M} and \citet{1980ApJ...237..303S} 
revealed the existence of shells or loops in elliptical galaxies. 
Then \citet{1988ApJ...328...88S} showed the presence of shell features in 
disk galaxies at a distance within $\sim$ 100 Mpc, suggesting that merging 
features are not exclusive to elliptical galaxies.
Furthermore, similar searches for merging features 
have been conducted for nearby galaxies (e.g., \citealp{2009AJ....138.1417T,2010AJ....140..962M,
2015MNRAS.446..120D,2015ApJ...800L...3W,2017ApJ...834...16M}) and more distant 
ones (e.g., \citealp{2005AJ....130.2647V,2012ApJS..202....8S,2013ApJ...765...28A}).

Recently, there have been several attempts to explore the properties 
of merging features and stellar halos in the outskirts of very nearby 
galaxies using the integrated light of galaxies 
(e.g., \citealp{2013ApJ...762...82M,2014ApJ...782L..24V,2016ApJ...830...62M,
2016ApJ...823..123T,2016ApJ...826...59W}), 
which requires not only deep imaging, but also wide fields-of-view (FoV) detector.
This method can be applicable to relatively distant galaxies 
and thus larger sample, compared to one of the optimal method using 
the resolved star counts in the stellar halos with Hubble Space Telescope 
(e.g., GHOSTS survey; \citealp{2011ApJS..195...18R,2017MNRAS.466.1491H}).
However, it is technically challenging to investigate photometric 
properties of the galaxy outskirts due to its diffuse structures 
with extremely low surface brightness. Therefore, it requires not only 
dedicated data reduction to recover the faint extended starlight 
from stellar halos, but also careful data analysis to quantify its 
physical properties. Despite these difficulties, they could 
detect merging features and stellar halos with surface 
brightness deeper than $\sim$ 28 mag arcsec$^{-2}$ and estimated 
their masses and colors. Previous studies \citep{2014ApJ...782L..24V,
2016ApJ...830...62M,2016ApJ...823..123T} have shown that the mass fraction 
of stellar halos substantially varies with a large scatter, but is in broad 
agreement with the predictions from the numerical simulations (e.g., 
\citealp{2010MNRAS.406..744C,2014MNRAS.444..237P, 2017MNRAS.466.1491H}). 
It reveals that the mass assembly history of galaxies might differ despite 
having similar total stellar masses.

\begin{figure}[t]
\centering
\includegraphics[width=85mm]{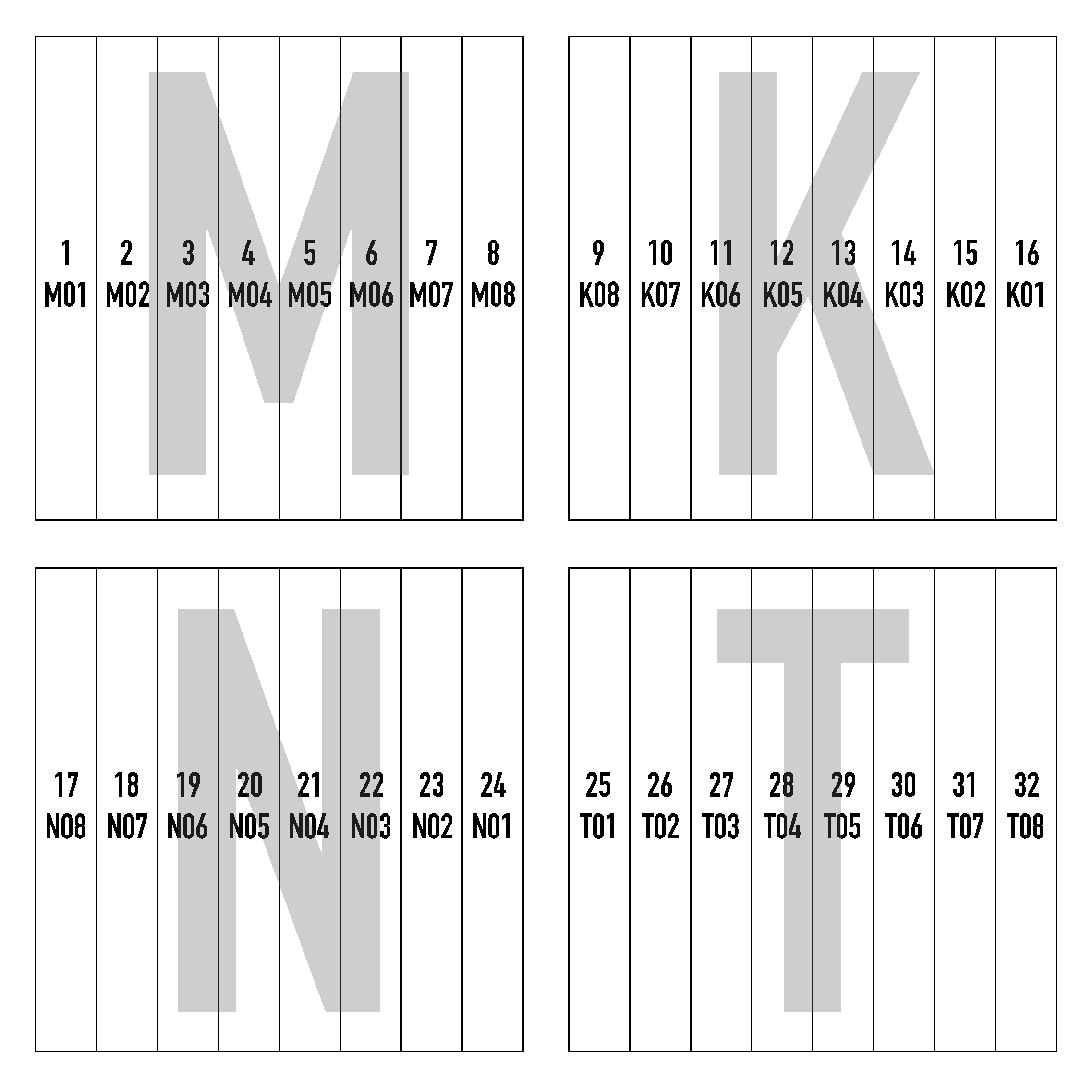}
\caption{The configuration of the KMTNet mosaic CCD imager. It consists of four 
9k$\times$9k CCD chips (M, K, N, and T), each of which employs eight 
amplifiers. The gap between CCDs is $\sim$ 184 arcsec in the 
horizontal direction and $\sim$ 373 arcsec in the vertical direction. 
The order and name of each amplifier are also indicated.\label{fig:fig1}
}
\end{figure}

In the same manner, we are conducting a deep imaging survey of nearby galaxies in the
southern hemisphere with Korea Microlensing Telescope Network 
(KMTNet; \citealp{2016JKAS...49...37K}).  KMTNet consists of three 1.6-m 
identical telescopes in three different sites (Cerro-Tololo in Chile, 
Sutherland in South Africa, and Siding Spring  Observatory in Australia). 
Each telescope contains
an imaging camera with a $2^{\circ}\times2^{\circ}$ of FoV to monitor 
stars in Galactic bulge in order to search for the microlensing events. 

Although KMTNet has wide FoV of detectors, it is not originally developed for detecting 
low surface brightness features. Therefore, an optimized observing 
strategy and data analysis with a great precision are required 
in order to discriminate the diffuse light in the galaxy outskirts from the 
scattered light, residuals of sky background and contamination from Galactic 
cirrus. 
To achieve the precision to reach a surface brightness of $\sim 27-28$ mag 
arcsec$^{-2}$, it is necessary to minimize the sky variation 
up to 0.5\% of the sky level through rigorous flat-fielding and sky subtraction. 
In that respect, the wide FoV of KMTNet detector enables us not only to determine sky values 
and global sky gradients precisely, but also to construct a dark sky flat with blank 
fields. However, no such attempts have yet been made in other KMTNet projects
\citep{2016JKAS...49...37K}. 
This experiment was the first attempt to overcome the difficulties 
caused by using sub-optimal systems for detecting integrated diffuse light in 
nearby galaxies. In this paper, we present observing strategy and optimal data 
reduction method to recover the extended faint structures using KMTNet 
data and describe initial results of our survey. Section 2 provides an overview 
of the observation strategy and targets. Section 3 describes the basic data reduction methods. 
Section 4 presents the sky determination method 
with imaging data of NGC 1291 and the radial profile of the galaxy to show how deep 
the survey data can reach, as a pilot study. We summarize our results in Section 5.

\section{Observations of NGC 1291}

A deep optical imaging survey of nearby galaxies was being conducted 
by taking advantage of the wide FoV ($2^{\circ}\times2^{\circ}$) of the three 
KMTNet 1.6-m telescopes. The KMTNet imager consisted of four 
9k$\times$9k CCD chips (M, K, N, and T), each of which employed 
eight amplifiers, as illustrated in Figure 1. 
The pixel scale is 0.4 arcsec and the gap between CCDs is $\sim$ 
184 arcsec in the horizontal direction and $\sim$ 373 arcsec 
in the vertical direction. 
The exposure time of each object frame is 120 seconds. 
Object frames were obtained by placing a target galaxy 
at the center of each chip to secure large sky areas for dark sky flat-fielding 
around a galaxy.
Then the images were taken following the 7-points dithering 
in each chip to fill the gap between four of the chips. 
As a result, the final stacked image would have 
a $3^{\circ}\times3^{\circ}$ FoV around the target galaxy. 
We obtained images of target galaxies for the same band using a 
single telescope at Cerro-Tololo to avoid troubles from different 
characteristics between three telescopes and keep consistency of dataset.

Our targets are originally selected from the catalog of 
Carnegie-Irvine Galaxy Survey (CGS; \citealp{2011ApJS..197...21H}), which conducted an imaging 
survey of nearby galaxies with $B_T \leq 12.9$ mag in the southern hemisphere.
As a pilot study, we chose NGC 1291 which has a largest angular size among
our sample, and also has a negligible contamination of Galactic cirrus. 
Therefore, this target is appropriate to test whether we can reach the 
precision required to study faint outskirts of nearby galaxies with KMTNet. 
Some basic properties of NGC 1291 are listed in Table 1. 
The data of NGC 1291 analyzed in this study were taken at KMTNet-CTIO observatory on 
November 12, 2015. During the night, the weather was clear and the average 
seeing was about 1.1 arcsec.

\begin{deluxetable}{lc}
\tablecolumns{2}
\tablewidth{0pc}
\tablecaption{Basic properties of NGC 1291\label{tab:jkastab1}}
\tablehead{
\colhead{Properties} &
\colhead{Value}
}
\startdata
Classification & (R)SB0/a(s)\\
Inclination & 12$^\circ$\\
R.A. (J2000) & 03$\mathrm{^h}$17$\mathrm{^m}$18$\mathrm{^s}$.6\\
Decl. (J2000) & -41$^\circ$06$^\prime$29$^{\prime\prime}$\\
Distance & 10.1 Mpc\\
$m_B$ & 9.44 mag
\enddata
\end{deluxetable}

\section{Data Processing}

\subsection{Overscan correction}

Prior to data reduction, we performed visual inspection of the dataset 
in order to discard images with artifacts due to mis-tracking of the 
telescope or abnormal jump of bias level in a part of detectors. 
Consequently, 85 $B$-band and 48 $R$-band images of NGC 1291 were 
utilized in the study, corresponds to the total exposure time of $\sim$ 2.8 and 
$\sim$ 1.6 hrs, respectively. We initially 
reduced the raw data using IRAF's {\tt mscred} package to process MEF 
(Multiple Extension FITS) files of the KMTNet.

First, overscan correction and image trimming were performed 
on all object and calibration frames. In general, it would be followed by 
bias correction with a master bias frame. 
However, we found that the bias levels had significantly varied for 10 
bias frames which were taken at the beginning and end of the observation.
In order to check the degree of variations in the bias level, we compared relative median 
counts of amplifiers along the different bias frames, as shown in the upper panel of Figure 2.  
It demonstrates that the bias levels varied up to 1\% in each amplifier 
during the night. Therefore, we decided not to conduct bias correction to the images because 
a master bias frame would not accurately represent the bias level of the data. 

According to our experiments, the bias levels could be corrected suitably when 
their own overscan regions were utilized. As shown in the bottom panel of 
Figure 2, we found that the bias levels were reduced to almost zero counts 
after overscan correction. Note that the original bias level is $\sim$ 1,700 counts on average.
Although there some offsets remained, they were less than 0.1\% 
of the original bias level, so their effect was negligible. 
The standard deviation of each bias frame was reduced to 
$\sim$ 0.26 counts, which is much smaller than the sky 
variation aimed for our deep imaging survey. 
From these results, we could assume that overscan is nearly synchronized with varying bias levels.
Therefore, we adopted overscan correction as a standard procedure for bias subtraction 
of the KMTNet data. Dark correction was not applied either because dark counts were 
negligible to be almost zero.

\begin{figure}[t]
\centering
\includegraphics[width=85mm]{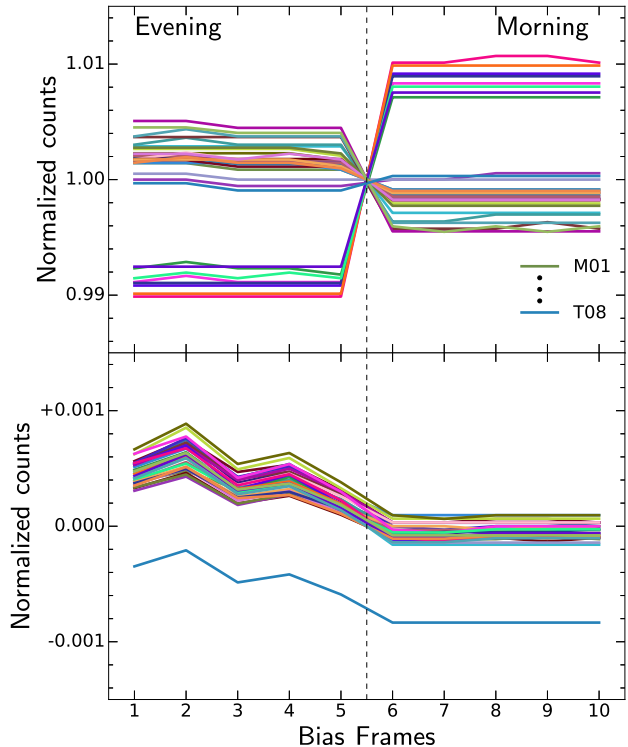}
\caption{{\it Top}: Variations of bias level in each amplifier along the 10 
bias frames taken in the evening and the morning of the observing run. 
Different colors represent different amplifiers. Counts are normalized to 
the median bias level of all amplifiers of 10 bias frames. 
Not only the bias levels 
significantly change during the night up to $\pm$ 1\%, but also 
their patterns are inconsistent with each other. 
{\it Bottom}: Same as the top panel, but after overscan correction is 
conducted. Although variations in bias level still remain, the amplitudes
of the variations are negligible relative to the original bias level ($<0.1\%$). Note that 
`T08' is unlike other amplifiers, but its residual is still less than 
$\sim$ 0.1\%.} \label{fig:fig2}
\end{figure}

\subsection{Dark sky flat-fielding}
KMTNet observatories did not provide dome flats in 2015, so we attempted
flat-field correction using twilight flats first. However, it turned out that 
it is almost impossible to keep a consistent illumination pattern between 
twilight flats over the large sky area of $2^{\circ} 
\times 2^{\circ}$ FoV. Figure 3 shows a comparison of twilight flats taken in the evening 
(solid line) and the morning (dashed line). The median counts for each 
amplifier were normalized by the median of all amplifiers in the same frame. 
We found that the illumination patterns of two twilight flats did not match, 
probably due to the combined effects of the intrinsic sky gradient varying over 
time and the wide FoV of the KMTNet CCDs.

Instead, we conducted dark sky flat-fielding for object frames. Dark sky flats 
were generated by stacking all object frames in each band. Prior to the 
stacking, all astronomical sources in the overscan-corrected object frames 
were masked using {\tt objmasks} in IRAF. The objects were initially 
identified as being 1.4$\sigma$ above the sky value. 
In addition, diffuse light near the target galaxy and artifacts around bright stars 
were manually masked out. 
The {\tt photutils} library in the PYTHON package was used 
to find the position and relative intensity of bright sources. The size of mask is enlarged 
according to the brightness of each masked source. In total, 85 $B$-band images and 
48 $R$-band images were combined for each band using {\tt sflatcombine} in 
IRAF to generate the master flat.

Figure 4 shows the object frames after twilight flat-fielding (left) and 
dark sky flat-fielding (right). 
We found that dark sky flat-fielding effectively removes inhomogeneous 
illumination patterns in object frames. 
The background variation among amplifiers was sufficiently eliminated and 
only intrinsic sky gradient remained across the image. 

\begin{figure}[t]
\centering
\includegraphics[width=85mm]{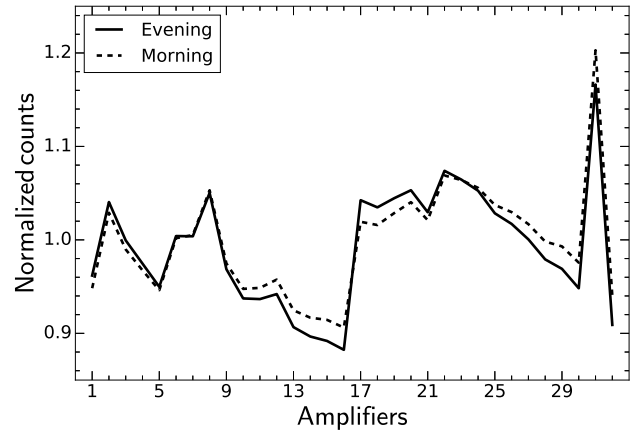}
\caption{
The normalized counts of the twilight flats in $R$-band as a function of 32 
amplifiers. The median value in each frame is normalized to the median of the 
all amplifiers in the same frame. Solid and dashed lines represent two 
twilight flats observed in the evening and morning, respectively. 
Due to the intrinsic sky gradient varying over time, the 
patterns between two twilight flats appear to change by up to
$\sim$ 5\% of the sky level. Note that twilight flats in $B$-band
exhibit the similar pattern.} \label{fig:fig3}
\end{figure}

\begin{figure}[t]
\centering
\includegraphics[width=85mm]{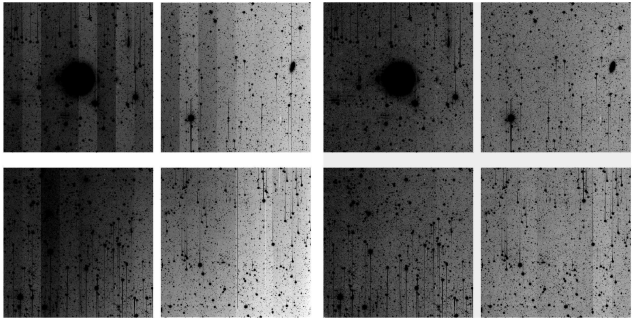}
\caption{The object frames after twilight flat-fielding (left) and dark sky 
flat-fielding (right). This difference  demonstrates that dark sky flat-fielding 
effectively corrects different background levels between amplifiers, creating a 
smooth background gradient better than twilight flat-fielding.}\label{fig:fig4}
\end{figure}

\subsection{Sky subtraction and Image combining}
We used the dark sky flat to generate the bad pixel mask that contains bad 
pixels, bad columns, and hot pixels. Those pixels were replaced by 
linearly interpolating the nearest pixels. 

After dark sky flat-fielding, there was still an intrinsic sky gradient 
arising from the combination of sky background distorted by zenith distance 
and the wide FoV of the KMTNet data as shown in the right image of Figure 4. 
We found that the shape and amplitude of the sky gradient varied significantly 
during the night. Peak-to-peak variation of the sky gradients
varied up to $\sim$ 3\% of the median sky value in both bands. Note that 
the peak-to-peak variation was defined as the fractional difference between 
the minimum and maximum sky values measured in each amplifier. 
This variation in sky gradients may have created global fluctuations in the 
backgrounds of stacked images, affecting the measurements of surface 
brightness radial profiles of target galaxies.

In order to remove the sky gradient from individual object images, we 
performed modeling of the two-dimensional sky in each object frame and 
subtracted the sky model from the image.
First, we merged 32 amplifiers in each object frame into a single 
extension image. Then we binned each object frame by $\sim 500 \times 500$ 
pixels, which is suitable for discarding unmasked hot pixels and smoothing over 
any peculiar features in the background. 
Then a two-dimensional polynomial fit with a second order for a sky background was 
conducted\footnote{The polynomial 
fits with higher orders do not appear to be superior to the second order fit.
Thus, we decided to adopt the second order fit in order to avoid removing the 
intrinsic faint features of the target galaxy during the sky 
subtraction (see also \citealp{2004ApJ...615..196F}).}. 
Figure 5 shows an example of the original sky gradient with object masks (left) 
and the sky model (middle) in the binned image. The blanked region shown in the 
left panel was not used for the polynomial fitting. It appears that the sky 
model with a two-dimensional polynomial fit worked reasonably well.  
The right panel of Figure 5 shows the sky-subtracted image using the two-dimensional 
sky model. The peak-to-peak variation was reduced to less than $\sim$ 1.65\% 
and $\sim$ 0.65\% of the original sky levels in $B$- and $R$- bands, 
respectively.

Before combining all object images into a mosaic image in each band, 
astrometric calibration was conducted following the instruction of
``Astrometric calibration for KMTNet data'',
provided by the KMTNet project team\footnote{\url{http://kmtnet.kasi.re.kr}}. 
The brief description of the process is as follows. 
1) We merged the amplifiers into four separate chips. 2) Using the 
coordinates of astronomical sources from each chip, we separately 
derived the astrometric solution using {\tt SCAMP} \citep{2006ASPC..351..112B}. 
3) The final astrometric solutions were applied to the sky-subtracted 
images. At last, the final mosaic images were generated using 
{\tt Swarp} \citep{2002ASPC..281..228B}.

\subsection{Standardization}
Photometric zero points for the combined images were determined using bright 
stars in the field of NGC 1291. We performed aperture photometry of the stars 
for the combined images using {\tt SExtractor} \citep{1996A&AS..117..393B} and compared 
them with the magnitudes in the catalog of AAVSO Photometric All-Sky Survey 
(APASS) DR9\footnote{\url{https://www.aavso.org/apass}}.
We chose the radius of aperture of 10 arcsec which is matched 
for that of APASS catalog. Note that APASS catalog provides 
the magnitudes of $B$, $V$, $g^{\prime}$, $r^{\prime}$, $i^{\prime}$ bands.
Therefore, we converted $g^{\prime}$ and $r^{\prime}$ 
magnitudes in APASS catalog to $R$ magnitude, using the conversion equation 
$R=r-0.1837(g-r)-0.0971$ provided by \citet{lupton05}. 

Figure 6 shows the comparison between the APASS magnitudes of stars and
their total counts from the KMTNet images measured within the aperture of $r =$ 10 
arcsec. We excluded saturated stars and ones in the crowded regions for the 
fit. The standardization equation for the combined images is as follows.
\begin{equation}
m_{*} = -2.5~{\rm log(count)} + m_{*,0} ~~~~ (*: B~or~R)
\end{equation}
The fit showed that the photometric zero points were determined to be 
$m_{B,0}\sim29.39\pm0.08$ and $m_{R,0}\sim29.52\pm0.06$.

\begin{figure}[t]
\centering
\includegraphics[width=85mm]{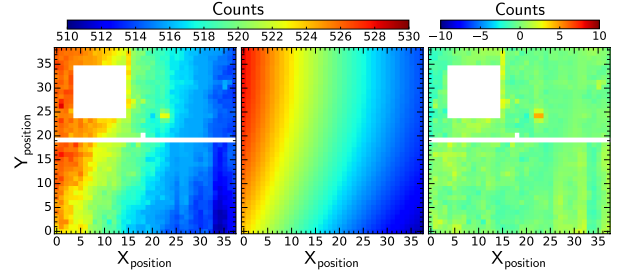}
\caption{{\it Left}: The intrinsic sky gradient of the right image in Figure 4, 
binned by $\sim 500 \times 500$ pixels. The target galaxy and the gap between 
chips are blanked. {\it Middle}: The sky value modeled by a two-dimensional polynomial 
fit with a second order. {\it Right}: The sky-subtracted image. 
The global pattern of the intrinsic sky gradient appears to be successfully
subtracted. The color represents the count of each pixel and the scale is 
shown on the upper side. \label{fig:fig5}}
\end{figure}

\section{Assessment: Radial Profile of NGC 1291}

In order to assess our deep optical images obtained with KMTNet, we derived
surface brightness radial profiles of NGC 1291 using the combined images.
First, we masked out all of the astronomical sources and then ascertained 
the sky level which should have been zero due to the optimal sky subtraction (\S{3.3}).
As shown in Figure 7, the medians of the sky level in the combined mosaic images are 
close to zero in both bands, and the 1-$\sigma$ variations of the sky backgrounds are 
0.8 DN/pixel and 2.1 DN/pixel in $B$ and $R$ bands, respectively. 
Therefore, the 1-$\sigma$ depths of surface brightness calculated based on 
the standard deviations
are $\sim$ 28.7 and 27.7 mag arcsec$^{-2}$ in $B$- and $R$-bands, respectively.
Distributions of sky values appears to follow Gaussian profile, revealing that the
background is uniform and the astronomical sources are properly masked out. 
Recently, \citet{2017ApJ...848...19P} reported newly discovered dwarf galaxies in NGC 2784 group 
using deep images obtained with KMTNet telescopes. The 1-$\sigma$ depths of surface brightness 
of their images are $\mu_B\sim$ 28.3, $\mu_V\sim$ 27.6, $\mu_I\sim$ 27.3 mag arcsec$^{-2}$, 
which are comparable with our results. It suggests that our data can be used to search 
dwarf galaxies around NGC 1291.

\begin{figure}[t]
\centering
\includegraphics[width=85mm]{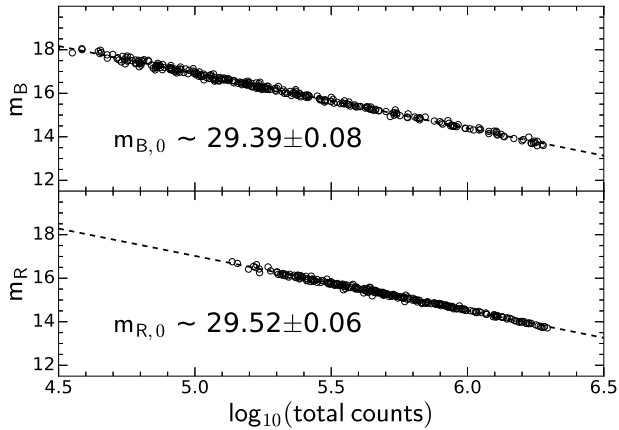}
\caption{The APASS magnitudes of the stars (circles) as a function of their total 
counts within an aperture ($r =$ 10 arcsec) in the combined images. Saturated 
stars and stars in the crowded regions were excluded from the fit. The results 
show tight correlations (dashed lines) between the magnitudes given by APASS 
and total counts measured in the images.}\label{fig:fig6}
\end{figure}

On the other hand, the 1-$\sigma$ depth in the azimuthally averaged 
one-dimensional profile can be significantly larger (fainter) than the values 
derived from the standard deviation of the sky background.
The overall uncertainty of mean surface brightness in each isophote can be reduced 
by $\sqrt{N}$ through the azimuthal average where $N$ is the number of pixels in the isophote. 
Because the target galaxy has a large apparent size, there can be a substantial number of pixels 
on each isophote in one-dimensional fit. Therefore, if the sky background is uniform, we could 
reach deeper surface brightness in one-dimensional profile  (see also \citealp{2013ApJ...762...82M,2016ApJ...830...62M}). 

\subsection{Sky determination}
Accurate sky determination is crucial to derive one-dimensional light profile precisely 
because the surface brightness of the outskirts of target galaxies is at least $50-100$ times 
fainter than the sky brightness. 
Although we carefully subtracted the sky value in the individual object frames, there can still 
remain residuals in the sky background. We found that the combined mosaic images exhibit 
a local spatial fluctuation around the target galaxy.  
The right panel of Figure 8 shows the distribution of sky levels in the vicinity of 
NGC 1291 in a $500\times500$ pixel-binned $R$-band image. The color represents the median count in each 
binned pixel. This sky fluctuation can introduce an uncertainty in the sky level measurement, 
which in turn can introduce additional error budget in the light profile.   
Sky fluctuations can be caused by imperfect masking of bright sources, and/or underlying 
Galactic cirrus, but neither can be regarded as the origin of the fluctuation. We found that 
the astronomical sources are properly masked from the distribution of the sky background and 
the sky fluctuation appears not to be coincident with nearby bright sources. 
From the IRAS map, we found $F_{100\mu m} \sim$ 1.5 MJy sr$^{-1}$ in the vicinity of 
NGC 1291, revealing the contribution from the Galactic cirrus is negligible (cf. \citealp{2016ApJ...830...62M}). 
While it is unclear what the origin of the the sky fluctuation is, we chose to take an 
uncertainty in the sky determination into account when we investigate the surface 
brightness profile of NGC 1291.

\begin{figure}[t]
\centering
\includegraphics[width=85mm]{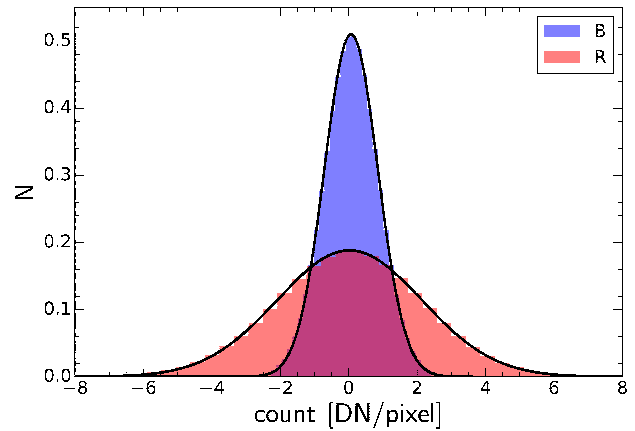}
\caption{The normalized histograms of the pixel values of sky background 
for combined images. Blue and red histograms represent $B$- and $R$-bands, 
respectively. Black lines show the fit with the Gaussian function.
Their peak positions are 0.07 and 0.02 DN/pixel and their standard deviations are 
0.8 and 2.1 DN/pixel in $B$- and $R$-bands, respectively.}\label{fig:fig7}
\end{figure}

To measure the sky level in the final mosaic image, we adopt four different regions. 
As shown in the left panel of Figure 8, we estimated sky values from different annuli along 
the distance from the galaxy center. \citet{2013ApJ...762...82M} presented that the starlight can be detected 
up to nearly $3 \times R_{25}$ in the deep images of M101. We conservatively chose the 
annulus for the sky determination beyond $\sim$ 20 arcmin, which is approximately 
equivalent to $4 \times R_{25}$ of NGC 1291 (cf. ${D}_{25}$ = 9.8 arcmin; 
\citealp{1975ApJS...29..193D}). The width of each sky region was set to $\sim 3.5$ 
arcmin. 
We found that the sky value varies between 0.02 (0.09) and -0.16 (0.06) for $R$($B$) band image, 
seemingly due to local sky fluctuations. 
Although these variations are less than $\sim 0.05\%$ of the original sky values, 
they can induce significant errors in surface brightness profiles, especially in the faint regions. 
In other words, our investigation for a deep imaging is mainly limited by local sky 
fluctuations, rather than 1-$\sigma$ variations in the sky background.
Once the sky level was estimated, we subtracted the sky value from the mosaic image.

\subsection{Surface brightness profiles of NGC 1291}
As shown in Figure 9, stellar light can be detected up to $\sim$10 arcmin in the two-dimensional 
image, which corresponds to  $R \sim$ 30 kpc. 
In order to investigate the capability of KMTNet for a deep imaging survey, 
we derived the one-dimensional surface brightness profile of NGC 1291 using the {\tt ellipse} 
task in IRAF. When conducting the isophotal fit, we fixed the central 
position but other parameters were allowed 
to be fit as free parameters. As we discussed in \S{4.1}, 
there was an uncertainty in the background level up to 0.05\% of the original 
sky values. In order to account for this uncertainty, we estimated the surface brightness 
profiles by varying sky levels as described in \S{4.1}. 
The best surface brightness profile was determined by employing the median value at a given 
radius. The scatter in the surface brightness profiles was implemented 
to compute the uncertainty. Note that the overall uncertainty in
the outskirts was dominated by the scatter due to the uncertainty in the 
background level rather than the variance within the isophote. 

\begin{figure*}[t]
\centering
\includegraphics[width=170mm]{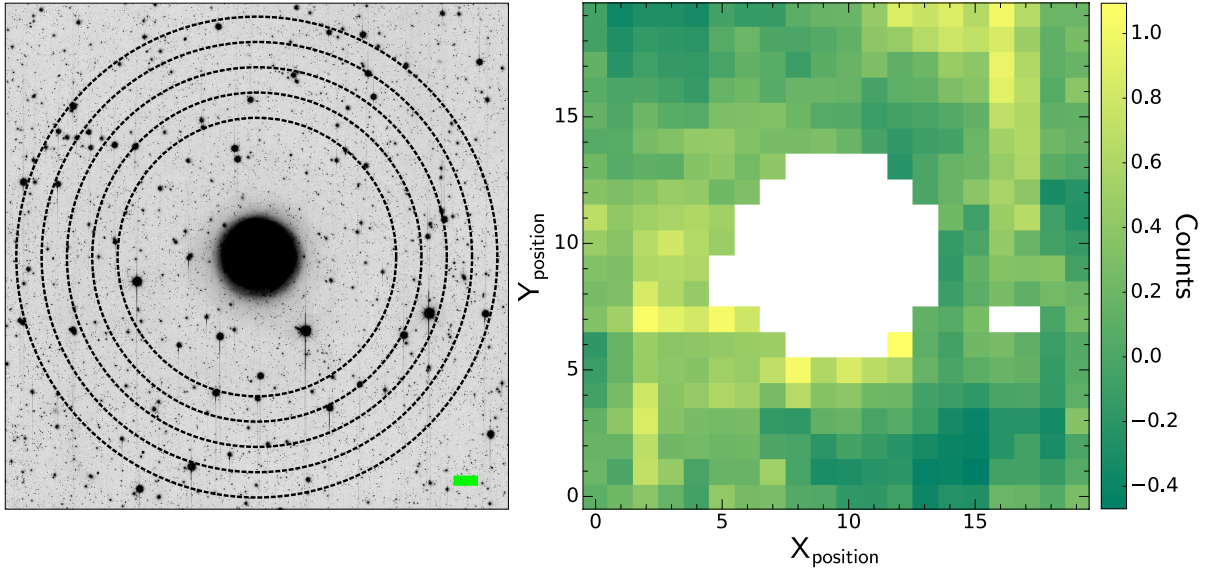}
\caption{{\it Left}: 
Four different regions used for sky determination are denoted as dashed 
lines. They are $20 - 30$ arcmin from the center of 
NGC 1291, and the width of the each region is 3.5 arcmin. The green horizontal 
bar indicates the scale of 3.5 arcmin ($\sim$ 10 kpc). {\it Right}: 
The distribution of sky levels in the $500\times500$ pixel-binned $R$-band image. 
The color represents the count of each pixel and the scale is shown 
on the right side. The FoV of the two panels is $\sim 1^{\circ}\times1^{\circ}$.}
 \label{fig:fig8}
\end{figure*}

Figure 10 shows the surface brightness profile of NGC 1291 in 
$B$- and $R$-bands. The 1$-\sigma$ depths of surface brightness 
in one-dimensional profiles are $\sim$ 29.5 and 28.5 mag arcsec$^{-2}$ 
in $B$- and $R$-bands, respectively. 
The vertically shaded area in Figure 10 denotes the inner radius 
($5\times R_h$; $20-25$ kpc) where the light from the stellar halo is 
expected to dominate the total stellar light \citep{2016ApJ...830...62M}. 
Theoretical and observational studies of the Milky Way Galaxy
have also suggested that stellar light is dominated by the halo rather than 
the disk component beyond a radius of $\sim$ 20 kpc
\citep{2010ApJ...712..692C,2014MNRAS.444..237P}.  
This is consistent with the fact that NGC 1291 and the Milky Way Galaxy have similar 
physical properties in terms of stellar mass 
($M_{*, {\rm NGC 1291}} \sim 8 \times 10^{10}$ \solmass; \citealp{2011ApJ...737...41L},
$M_{*, {\rm MW}} \sim 4-5 \times 10^{10}$ \solmass; \citealp{2013ApJ...779..115B}) 
and half-light radius ($R_{h, {\rm NGC1291}}\sim 4-5$ kpc, 
$R_{h, {\rm MW}}\sim 2.74$ kpc; \citealp{2010AJ....140.1043V}).
It reveals that our deep images obtained with KMTNet are deep enough to
detect relatively bright features such as tidal streams, outer disks and
dwarf companions (cf. \citealp
{2010AJ....140..962M}), but it might be challenging but not 
impossible to detect diffuse stellar halos at the outskirts 
of galaxies \citep{2008ApJ...689..936J,2011ApJ...739...20C,2012arXiv1204.3082B}.
For this reason, our ongoing survey will focus on the target of extended
UV-disk (XUV-disk) galaxies which tend to have brighter and more discrete
features at the outskirts than diffuse stellar halos \citep{2007ApJS..173..538T}.

Although investigating the detailed structural properties is beyond the scope 
of this study, we briefly describe characteristics of the surface brightness 
profile of NGC 1291. NGC 1291 is known as a barred galaxy with an outer ring
with a radius of $\sim$ 9 kpc. The outer ring is more prominent in both UV and FIR
images than in optical images, suggesting that a significant number of young 
stars and amount dust reside in the ring \citep{2012ApJ...756...75H}.
The feature of the outer ring can be clearly seen in the one-dimensional profile in 
Figure 10 (black arrow). While the color of the profile appears to be 
approximately constant in $2-25$ kpc, $B-R$ color is the smallest (bluest) 
around the region of the ring, which is consistent with previous studies.
The disk in spiral galaxies can be classified as three different groups 
depending on the shape of the disk profile : pure exponential profile (Type I);
truncation profile (Type II); and anti-truncated profile (Type III); 	
\citep{2006A&A...454..759P}. 
Applying the imaging decomposition for 1-D profile of NGC 1291, we found that
the disk is well described by a single exponential profile, indicating that
NGC 1291 has Type I disk.
The lack of dramatic variation in $B-R$ color is 
also consistent with color profile of Type I disk \citep{2008ApJ...683L.103B}.

One of the ultimate goals for this survey is to explore the halo structures 
using deep KMTNet images. \citet{2016ApJ...830...62M} suggested that the halo
features can be prominent at a radius above $5\times R_h$, which is equivalent
to $20-25$ kpc for NGC 1291. Figure 10 shows that the light profile beyond 
$20-25$ kpc does not deviate from the extrapolation of the exponential profile 
derived from the inner part of $R<25$ kpc. 
It reveals that the halo contribution above a surface brightness of 
$28-29$ mag arcsec$^{-2}$ is insignificant in NGC 1291. Note that one would 
expect a few percent of stellar halo fractions at this mass 
(e.g., \citealp{2011ApJ...739...20C,2016MNRAS.458..425V}). 
This finding is in broad agreement with the results of
Merritt et al. (2016), in the sense that spiral galaxies have a wide range
of stellar halo mass fractions. 

\begin{figure}[t]
\centering
\vspace{-2.5mm}
\includegraphics[width=85mm]{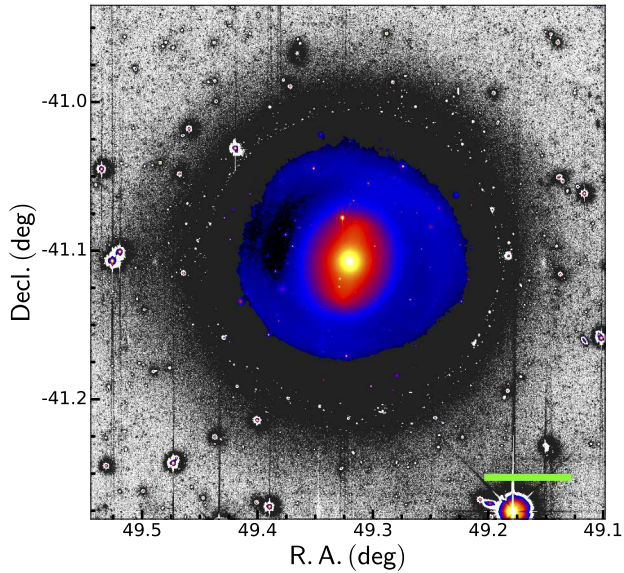}
\caption{The combined $R$-band mosaic image of NGC 1291. 
While the faint extended region is shown in grayscale, 
the central part with the brighter surface brightness level is color-coded. 
The green horizontal bar indicates the scale of 10 kpc at a distance of 10.1 Mpc.} 
\label{fig:fig9}
\end{figure}

\section{Summary}

We are carrying out a deep wide-field imaging survey of nearby galaxies in the
southern hemisphere using KMTNet. The aim of this survey is to investigate the faint and 
extended structures of galaxies which contains the mass assembly histories of galaxies. 
In this study, we assess if imaging dataset of KMTNet is suitable to achieve our goal. 
We describe the detailed data reduction process using images of NGC 1291, 
one of the largest of the target galaxies.
The data reduction was conducted as follows.
\begin{itemize}
\item The bias level appeared to vary up to $\pm$1\% over time, so we 
adopt overscan correction instead of using the master bias for the bias 
correction. The uncertainty due to the variation of the bias level was 
significantly reduced below $\sim$ 0.015\% to the original bias level.
\item The twilight flat frames of KMTNet were severely affected by 
the intrinsic sky gradient owing to their wide FoV. 
Therefore, the twilight flats appeared to be inappropriate for flat-field correction. 
Instead, we generated the dark sky flat using the object frames for flat-fielding. 
After flat-field correction with the dark sky flat, a peak-to-peak variation 
among amplifiers can be reduced up to $\sim$ 3\% of the original sky levels, 
which originated in the intrinsic sky gradient in the object frames.
\item To remove the intrinsic sky gradient of the object frames, we 
employed a two-dimensional polynomial function with a second order to model the sky background, 
which is subtracted from the object frames. In so doing, the peak-to-peak variation 
among amplifiers was reduced to less than $\sim$ 1.65\% and $\sim$ 0.65\% of the original 
sky levels in $B$- and $R$-bands, respectively.
\end{itemize}

\begin{figure}[t]
\centering
\includegraphics[width=85mm]{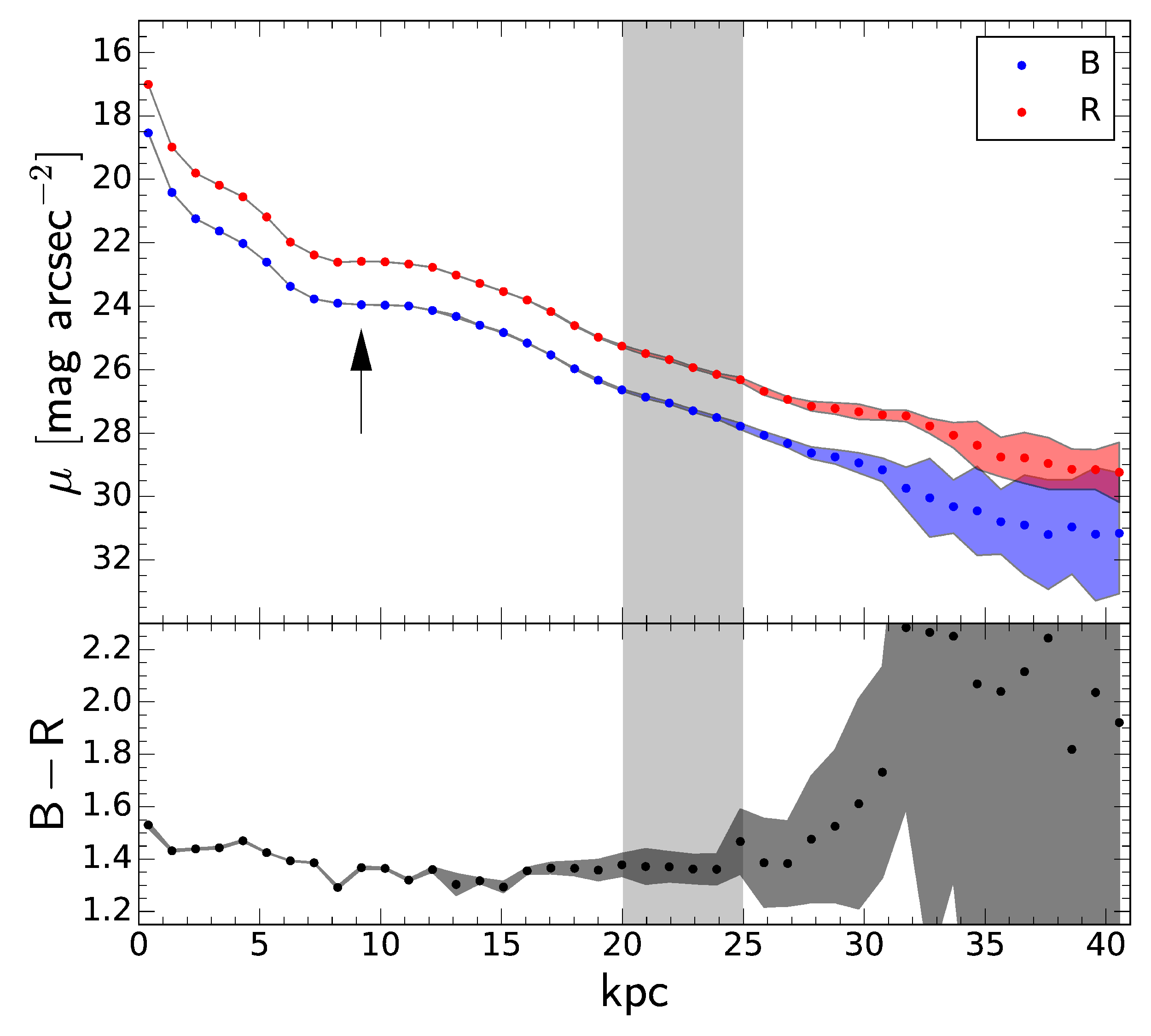}
\caption{{\it Top}: The surface brightness profiles of NGC 
1291. The galaxy outskirts appear to be detected up to $\sim$ 34 kpc with 
a 1-$\sigma$ depth of $\sim 28.5-29.5$ mag arcsec$^{-2}$. 
The error budget (lightly shaded region) is estimated to be the quadratic sum of  
the variance in the isophote and the scatter of the profiles induced by the 
uncertainty in sky determination. The position of the outer ring is indicated 
by an arrow. 
The vertically shaded area represents the inner radius where the stellar halos 
are expected to dominate the stellar light (see the text for details).
{\it Bottom}: $B-R$ color profile. 
It appears to have a nearly flat gradient up to $\sim$ 25 kpc. 
} \label{fig:fig10}
\end{figure}

Finally, we generated the combined mosaic images with FoV of
$\sim3^{\circ}\times3^{\circ}$ in $B$- and $R$-bands. 
We found that a local spatial fluctuation exists in the vicinity of the 
target galaxy, inducing the uncertainty in sky determination.
Although the variation of the sky level due to the sky fluctuation was less than 
$\sim$ 0.05\% of the original sky brightness, it introduced an additional and 
dominant uncertainty in the surface brightness profile. Taking this 
effect into account, we derived the surface brightness profiles of NGC 1291. 
As a result, we were able to reach the surface brightness of $\mu_{B,1\sigma}\sim$ 
29.5 and $\mu_{R,1\sigma}\sim$ 28.5 mag arcsec$^{-2}$ in 
one-dimensional light profiles, respectively. 
It suggested that the deep images obtained with KMTNet are suitable to 
detect the tidal debris, outer disks and dwarf galaxies 
\citep{2010AJ....140..962M}, but might not be ideal to analyze faint diffuse 
stellar halos \citep{2008ApJ...689..936J,2011ApJ...739...20C,
2012arXiv1204.3082B}.  
As we discussed above, the achieved imaging depth is mainly limited by
the local fluctuation of the sky value rather than the integration time or
large-scale sky variation. Therefore, our observing strategy is well
optimized to achieve our scientific goal.
Based on the result of this study, we decided to focus on exploring outer
disks in XUV-disk galaxies, and dwarf companions in nearby late-type galaxies,
which are thought be brighter than our depth limit. In addition, we will also
obtain $I-$band and H$\alpha$ imaging data for better understandings of the 
stellar populations and current star formation in the outskirts.
We are planning to expand sample size and 
investigate photometric characteristics of the outskirts of individual target 
galaxies including extended disk profiles and dwarf companions in upcoming papers.

\vskip 0.3in


We are grateful to an anonymous referee for constructive comments and suggestions.
We thank Zhao-Yu Li and Hua Gao for useful discussion. 
This research has made use of the KMTNet system operated by the Korea 
Astronomy and Space Science Institute (KASI) and the data were obtained at 
three host sites of CTIO in Chile, SAAO in South Africa, and SSO in Australia.
This research was supported by the Basic Science Research Program through the
National Research Foundation of Korea (NRF) funded by the Ministry of
Science, ICT \& Future Planning (No. NRF-2017R1C1B2002879). L.C.H. was supported by the 
National Key R\&D Program of China (2016YFA0400702) and the National Science Foundation 
of China (11473002, 11721303).

\end{document}